\documentstyle[12pt,fleqn]{article}
\def\beqa{\begin{eqnarray}}
\def\eeqa{\end{eqnarray}}
\def\beq{\begin{equation}}
\def\eeq{\end{equation}}

\def\vol{\int d^4x\,\sqrt{-g}}

\def\half{\frac{1}{2}}

\def\umu{^{\mu}}

\def\dmu{_{\mu}}

\def\umunu{^{\mu\nu}}
\def\dmunu{_{\mu\nu}}

\def\dab{_{\alpha\beta}}

\def\udeab{^{;\alpha\beta}}

\def\ddemunu{_{;\mu\nu}}

\def\ddemu{_{;\mu}}  
\def\ddenu{_{;\nu}}  
\def\ddea{_{;\alpha}}  \def\udea{^{;\alpha}}
  \def\udeb{^{;\beta}}

\def\p{\partial}
\def\bib#1{$^{\ref{#1}}$}
\let\lam=\lambda

\let\Gam=\Gamma

\let\ome=\omega

\def\rpr{ Phys. Rev. }
\def\rprl{ Phys. Rev. Lett. }
\def\rpl{ Phys. Lett. }
\def\rnp{ Nucl. Phys. }

\def\rijmp{ Int. J. Mod. Phys. }

\def\rcqg{ Class. Quantum Gravit. }

\def\rgrg{ Gen. Relativ. Gravit. }

\def\rapj{ Ap. J. }
\def\rapjl{ Ap. J. Lett. }

\begin{document}
\begin{flushright}
gr-qc/9302010
\end{flushright}
\vspace{.3in}
\centerline{\large \bf Cosmology with Nonminimal Derivative Couplings}
\vspace{.3in}
\centerline{LUCA AMENDOLA}
\vspace{.2in}
\centerline{{\it Osservatorio Astronomico di Roma}}
\centerline{\it Viale del Parco Mellini, 84}
\centerline{\it I-00136 Rome, Italy}
\vspace{.2in}
\centerline{ABSTRACT}
\baselineskip 14pt
\begin{quote}
We study a theory which generalizes the nonminimal coupling
of matter to gravity by including derivative couplings. This leads to
several interesting new dynamical phenomena in cosmology. In particular,
the range of parameters in which inflationary attractors exist is greatly
expanded. We also numerically integrate the field equations and draw
the phase space of the model in second order approximation.
The model introduced here may display different inflationary epochs,
generating a non-scale-invariant fluctuation spectrum without the
need of two or more fields.
Finally, we comment on the bubble spectrum arising during a first-order
phase transition occurring in our model.
\end{quote}
\vspace {.1in}
\baselineskip 14pt
\normalsize

\section{\normalsize\bf Introduction}
Scalar fields in General Relativity has been a topic of great
interest in the latest years, mainly because the scalar field
dynamics allows to investigate
the detailed features of the early Universe. Without
the need of a specific equation of state, a Universe filled by
a phenomenological scalar field leads to an accelerated phase of
expansion, the inflation, in almost any kind of self-interaction
potential. As it has been shown in chaotic models of inflation\bib{LIN},
the accelerated phase, either power-law\bib{LM} or quasi-exponential, is
a phase-space attractor for most of the initial conditions, and
this result also extends to a large class of inhomogeneous and
anisotropic space-times\bib{WM}. The class of successful models has been
enlarged to scalar fields with a nonminimal coupling (NMC) to
the curvature scalar $R$ in the gravity Lagrangian\bib{KFM},
 commonly in the form
of a term $\sqrt{-g}f(\phi) R$, where $f(\phi)$ is a function of the
scalar field $\phi$. The motivations for
this step are manyfold: the idea that the fundamental
constants are not constant, the ``machian'' theory of gravitation
embodied in the Jordan-Brans-Dicke theory\bib{JBD},
 the renormalizing term arising in quantum field theory in curved
space\bib{BD},
the possibility to have gravity as a
spontaneous symmetry-breaking effect\bib{ZA}, the Kaluza-Klein
compactification scheme\bib{KT}, the  low-energy limit
of the superstring theory\bib{SS}.
Moreover, the
NMC term has been employed to produce an oscillating Universe,\bib{MM}
to reconcile cosmic strings production with inflation,\bib{YOK} to generate
a modified Newtonian dynamics able to model flat  rotation curves
in galaxies.\bib{BM} Last but not least,
an NMC term allows to solve the ``graceful exit'' problem of the old inflation
by slowing the false-vacuum expansion, as in extended inflation\bib{LS}.
Recently, the NMC models have been generalized in various directions:
models with different coupling functions\bib{SA}, with generalized coupling
to the inflaton sector\bib{HK}, with fourth-order gravity\bib{ACLO},
and with a coupling to a dark matter sector\bib{DGG}.

In this Letter we wish to further explore the influence of
nonminimal couplings in cosmology by introducing the nonminimal
derivative coupling (NMDC) to gravity. In this class of models, the
coupling function is also function of the derivatives of $\phi$:
$f=f(\phi,\phi\ddemu,\phi\ddemunu,...)$. Derivative couplings,
although rares, are not a novelty in physics: the field
theory of scalar quantum electro-dynamics includes derivative
couplings between the electro-magnetic vector $A\dmu$ and the
scalar field $\phi$; this kind of interaction is indeed
required by the $U(1)$ gauge-invariance of the theory.
In addition, if the idea underlying the
nonminimal coupling theory is that the Newton
constant $G$ itself depends on the gravitational field source mass,
it seems more natural to couple the curvature to the
energy-momentum tensor of the matter, introducing the
terms $TR$ and $T\dmunu R\umunu$, which in fact contain derivative
couplings. The NMDC model can be thought of as the scalar field
formulation of the hypothesis that $G=G(\rho)$, where $\rho$ is
the energy density of the gravitational field source.

One of the aims of this
Letter is to see whether the presence of NMDCs allows to find  inflationary
attractors in models where otherwise they are not present.
We studied a large class of NMC models in Ref. (\ref{ABO}). For future
reference, let us summarize the main results of Ref. (\ref{ABO}).
Hereinafter, we will use $f=f(\phi)$ referring to the non-derivative terms
coupled to the curvature scalar,
and $f_1=f_1(\phi,\phi\ddea)$ when we
refer to all the terms, including the
derivative terms, coupled to the curvature scalar.
In Ref. (\ref{ABO}) the coupling function $f(\phi)$ is left as general
as possible, and the
potential is assumed to be $V=\lam f^M$. It is also assumed that
$f(\phi)$ is semi-positive definite and that, for large $|\phi|$,
it grows monotonically faster than $\phi^2$. The case $f\sim\phi^2$ is
the standard choice, and has been extensively treated in
the literature\bib{KFM}.
 The relation between the potential and the coupling function
assumed in Ref. (\ref{ABO})
allows to determine the class of models with successful inflationary
attractors; here ``successful'' means that the inflation eventually stops
and a Friedmann stage begins. In Ref. (\ref{ABO}) and in the present work the
inflationary attractors are all asymptotic, in the limit of large
$|\phi|$; in this limit the influence of the coupling is clearly larger.
Moreover, the spirit of chaotic inflation is that whatever the initial
conditions of the Universe are, the dynamical trajectory
falls onto an inflationary attractor. This is verified only if the attractor
itself extends to very large values of the field variables, until the
classical description loses meaning (i.e. near the Planck boundary).
In Ref. (\ref{ABO}) it is shown that such asymptotic attractors exist only
if $2\le M<2+\sqrt{3}$, whatever functional form of $f(\phi)$
is considered. In this range, it is found for the cosmic scale factor $a(t)$
a power-law inflation $a\sim t^p$ with exponent
\beq\label{NMC}
p={3+(M-2)^3\over (M-2)^2(M-1)}\,.
\eeq
In the conformally rescaled frame the power-law exponent has
a much simpler form, $\tilde p= 3/(M-2)^2$. If
$M=2$ a deSitter exponential inflation is found. Further, in the cases
in which a successful inflation exists, the spectrum of primordial
fluctuations is independent of $f(\phi)$ and depends on
the wavenumber $k$ according to the power-law
$k^{1/(1-\tilde p)}$. Finally, in Ref. (\ref{ABO}) it is shown that the
fluctuations have a Gaussian distribution.
In the present work we extend some of the cited results to
derivative couplings. We assume as in Ref. (\ref{ABO}) that $V=\lam f^M$,
but now we will freeze the functional form of $f(\phi)$ to a power-law,
$f(\phi)=\phi^{2m}$. We will show that the narrow range in which
the parameter $M$ is confined in NMC models is greatly expanded when
one includes a derivative coupling.

It rises natural the question
of whether the conformal
transformation\bib{W}, which allows one to put complicated unconventional
gravity theory in pure Einsteinian form with one
or more scalar fields, works also when derivative couplings are
present. We will show schematically that a conformal
rescaling $\tilde g\dmunu=e^{2\omega} g\dmunu$ cannot recast
our theory in Einsteinian form. In order to recover
the Einstein field equations
the metric transformation should be generalized to a Legendre
transformation, as in Ref. (\ref{MFF}). This would introduce in
the equations additional tensor fields, instead
of additional scalar fields as in non-derivative cases.
Due to this reason, we avoided here the use of metric transformations.

The plan of this paper is as follows: in the next section we introduce
the derivative couplings, in Sect. 3 we
derive the equations of motion for the fields,
in Sect. 4 we perform the second-order approximation,
in Sect. 5 we discuss the phase space of some models
and finally in Sect. 6 we present
the conclusions.

\section{\normalsize\bf Derivative couplings}
Terms with NMDCs generalize the Einstein-Hilbert Lagrangian.
The most general gravity Lagrangian linear in the curvature scalar
$R$, quadratic in $\phi$, containing terms with four
derivatives includes all of the following terms:
\beqa\label{terms}
L_1&=&\mu\phi\ddea\phi\udea R\,;\quad L_2 = \tau\phi\udea\phi\udeb R\dab;
 \quad L_3=\eta\phi\Box\phi R \,;\nonumber\\
L_4 &=&\theta \phi\phi\udeab R\dab\,;\quad
L_5= \nu\phi\phi\ddea R\udea; \quad L_6 =\sigma\phi^2 \Box R\,.
\eeqa
The constants $\mu,\tau,...$ have the same dimensions in natural units
as the Newton constant $G$, namely $~mass^{-2}$.
Here and in the following the conventions are:
signature $\,(+---)\,$; units $\, 8\pi G=c=1\,.$
Due to the following relations
\beqa\label{div}
0&=&\vol \Box(\phi^2 R)=\vol\left[2\phi\Box\phi R+ 2\phi\ddea\phi\udea
R+\phi^2\Box R\right]\,,\nonumber\\
0&=&\vol\nabla\dmu(R\nabla\umu\phi^2)=\vol\left[2\phi\phi\udea R\ddea+
2\phi\Box\phi R+2\phi\ddea\phi\udea R\right]\,,\nonumber\\
0&=&\vol\nabla\dmu(\phi\phi\ddenu R\umunu)=\vol\left[\phi\ddemu\phi\ddenu
R\umunu+\phi\phi\ddemunu R\umunu+\phi\phi\ddenu R\umunu\ddemu\right],
\eeqa
three of the six terms $L_1-L_6$ can be neglected (also taking into
account the Bianchi identities). Other total divergences,
like $(\phi^2 R\umunu\ddenu)\ddemu$ and $(\phi^2 R\umunu)\ddemunu$
are linear combinations of Eqs. (\ref{div}).
Nevertheless, the
equations remain hopelessly complicated. We take here a first step.
We keep only the derivative term $L_1$, together with
the NMC correction $L_0=\xi f(\phi) R$. This particular choice
turns out to be the simplest one. Moreover, the terms studied here
are those appearing in the simplest ``natural'' coupling, namely $TR$.
In this way we
may draw an useful comparison between Ref. (\ref{ABO}), where
$L_0$ was the only correction to Einstein gravity, and the present
work.
We are then left with the following Lagrangian density
\beqa\label{lagr1}
L&=&-R+f_1(\phi,\phi\ddea) R+\phi\ddea\phi\udea-2 V(\phi)\,,\nonumber\\
f_1(\phi,\phi\ddea) &=& \xi f(\phi)+\mu\phi\ddea\phi\udea\,.
\eeqa

\def\dudemunu{_{1;\mu\nu}}
\section{\normalsize\bf Field equations }
The set of equations derived from (\ref{lagr1}) is
\beqa\label{einst}
G\dmunu (1-f_1) &=& g\dmunu\Box f_1-f\dudemunu+
\mu\phi\ddemu\phi\ddenu R+T\dmunu\,,\nonumber\\
T\dmunu &=&\phi\ddemu\phi\ddenu-\half g\dmunu\phi\ddea\phi\udea
+V(\phi) g\dmunu\,,
\eeqa
while the scalar field equation is
\beq\label{phi}
\Box\phi(1+\mu R)+\mu
\phi\ddea R\udea-\xi f' R/2+V'=0\,.
\eeq
It can already be seen that, even if the new terms introduced by
the derivative coupling (let us call them $\mu$-terms) do not determine
the critical points of the theory, they do contribute to the asymptotic
dynamical properties of Eq. (\ref{phi}). Indeed we will see that
 in the limit of large $R$ and $R\ddea$ (in other words, in the
early Universe), the $\mu$-terms will
sensibly modify the behavior of the non-derivative theory.

The trace of Eqs. (\ref{einst}) reads simply
\beq\label{ricci}
 R(\xi f-1)= T+3\Box f_1\,,
\eeq
where $T=-\phi\ddea\phi\udea+4V$.
The derivative of the Ricci scalar, to be used in Eq.  (\ref{phi}), is
\beq\label{riccider}
R\ddea={1\over\xi f-1}\left[T\ddea+3\Box f_{1,\alpha}-\xi f'\phi\ddea
R\right]\,.
\eeq
In a Friedmann-Robertson-Walker (FRW) spatially flat
metric with scale factor $a(t)$,
the ($0,0$) component of (\ref{einst}) is
\beq\label{00}
H^2(1-f_1)=H\dot f_1+{\mu\over 3}R\dot\phi^2+
{1\over 3}\left(\half\dot\phi^2+V\right)\,,
\eeq
where $H=\dot a/a$ and
where our conventions are such that $R=-6\dot H-12 H^2$. From now on,
we confine ourselves to a FRW spatially flat metric.
The system (\ref{phi}-\ref{00}) is closed. We have
five degrees of freedom $H,\phi,\dot\phi,\ddot \phi,\phi^{III}$
and indeed we have one equations of fourth order in $\phi$
[Eq. (\ref{phi})] and one of first order in $H$ [Eq. (\ref{00})].
The solution of this system appears very difficult. In the next
section we will make use of a second-order approximation
to derive the asymptotic properties of our model and to sketch its
phase space.

Let us conclude this section showing how a conformal
transformation operates on the NMDC theory.
Ordinary NMC theories are recast in Einstein form by
the metric rescaling\bib{W}
$\tilde g\dmunu=e^{2\ome}g\dmunu$, with
\beq\label{cf}
e^{2\ome}=|\p L/\p R|\,.
\eeq
Adopting the same formula, in our case we get $e^{2\ome}=|1-f_1|$.
Now, we have that the Einstein tensor $G\dmunu=R\dmunu
-Rg\dmunu/2$ trasforms according to
\beq\label{g}
G\dmunu=\tilde G\dmunu-2\left[\ome\ddemu\ome\ddenu+\half g\dmunu\ome\ddea
\ome\udea+g\dmunu\Box\ome-\ome\ddemunu\right]\,.
\eeq
Inserting (\ref{g}) into (\ref{einst}), with the choice (\ref{cf}), one has
\beq\label{einsttc}
\tilde G\dmunu=3 e^{-4\ome}\left( f_{1;\mu} f_{1;\nu}-\half g\dmunu
f_{1;\alpha} f_1\udea\right)+\mu e^{-2\ome}\phi\ddemu\phi\ddenu R
+e^{-2\ome}T\dmunu
\,,\eeq
which is {\it not} in Einsteinian form.
Any other choice of the conformal factor would not cancel the
second derivatives of $f_1$ which are present in (\ref{einst}).
Then, the right-hand-side of (\ref{einsttc}) could not be
written as a scalar field energy-momentum tensor.
Notice that $R$ at the right-hand-side is  not
trasformed. If one also
trasforms $R$ in (\ref{einsttc}), additional terms like $(\Box\ome)\phi\ddemu
\phi\ddenu$ arise.

\section{\normalsize\bf Second-order approximation}
In slow-rolling motion one neglects all terms with more than one time
derivative in the field equations. In this section we are more general, and
we examine the field equations by neglecting all terms of order
higher than the second one. This gives us three advantages: the
equations are notably simplified, we can calculate analytically some
inflationary solutions, and we can plot the phase-space portrait
of the model. Neglecting all terms higher than second order in (\ref{phi})
we have
\beq\label{phis}
\Box\phi(1+\mu R^{(0)})+\mu
\dot\phi \dot R^{(1)}-\xi f' R^{(2)}/2+V'=0\,.
\eeq
The indexes in parentheses denote the order to which $R$ and $\dot R$
are to be calculated. From Eqs. (\ref{ricci},\ref{riccider})
 and for $\xi f\gg 1$ we have
\beqa\label{ricci2}
R^{(0)} &=& {4V\over \xi f}\,,\\
R^{(2)} &=& {1\over \xi f}\left[-\dot\phi^2+4V
+3\xi\Box f)\right]\,,\\
\dot R^{(1)} &=& {4\dot\phi\over \xi f^2}\left[V'f-Vf'\right]\,.
\eeqa
We now explicitely adopt the functional relation $V=\lam f^M$. This
will simplify our equations. Eq. (\ref{phis}) then writes
\beqa\label{inse}
&&\Box\phi\left[f+{4\mu\lam\over\xi}f^M-{3\over 2}\xi f'^2\right]
+f'\dot\phi^2\left[{4\mu\lam\over\xi}f^{M-1}(M-1)+\half (1-3\xi f'')\right]
\nonumber\\
&&+\lam f' f^M (M-2)=0\,.
\eeqa
Let us call $A$ the coefficient of $\Box\phi$ and $B$ the coefficient of
$f'\dot\phi^2$. In order to avoid singularities in Eq. (\ref{inse})
one should have a nonvanishing coefficient $A$ for $\xi f\gg 1$
(in the opposite limit the equations reduce to the minimally coupled
model and the regularity follows automatically).
A sufficient
condition, although not a necessary one, is that $\xi,\mu\le 0$,
along with the already imposed condition $f\ge0$.
This assumption greatly simplifies the investigation
of the dynamical properties of the model under study and will
be adopted in the following. It also makes clearer the comparison with
previous work in ordinary NMC theories, where the condition
$\xi<0$ is often adopted to get a regular metric rescaling.
Notice that $\xi,\mu\le0$ ensures the positivity of the effective
Newton constant in the Lagrangian density (\ref{lagr1}).

The d'Alambertian operator reads $\Box\phi=\ddot\phi+3H\dot\phi$
in FRW metric, where the Hubble function $H$ is to
be taken from Eq. (\ref{00}) (to the first order):
\beq\label{00con}
\xi H^2 f+\xi H\dot f+{4\mu V\over3\xi f}\dot\phi^2+
{1\over 3}\left(\half\dot\phi^2+V\right)=0\,.
\eeq

The asymptotic behavior of Eq. (\ref{inse}) can be now easily discussed in
the limit in which some of the terms in $A$ and $B$ can be neglected.
A first consideration is however immediate: for $M<2$ the second
derivative of the effective potential in Eq. (\ref{inse}) is
negative, in the hypothesis that $f$ grows monotonically with
$|\phi|$. As a consequence, any attractor solution of Eq. (\ref{inse})
leads to ever growing $f$; the Friedmann behavior cannot
be reached. Let us make an example.
If we assume $f(\phi)\to \phi^{2m}$ for large $|\phi|$, it turns out
that if $M<2-1/m$ one has that $f'^2$ dominates in $A$ and $f''$ in $B$. Then
Eq. (\ref{inse}) simplifies to
\beq\label{fcase}
\Box f-[2\lam (M-2)/3\xi] f^M=0
\eeq
(where $\Box f=f'\Box \phi+f''\dot\phi^2$). In this case the influence of
the $\mu$-terms disappears; the model reduces to the ordinary NMC theory,
already discussed in Ref. (\ref{ABO}). There it was found that for $M<2$ no
successful inflation was allowed. Indeed, in Eq. (\ref{fcase}) the
effective potential derivative is negative definite (since $\xi<0$ and $M<2$).
As already discussed, the attractor solutions of  Eq. (\ref{fcase}) are
directed
toward increasing $f$; the Friedmann behavior will never be recovered.
The same conclusion holds
for any $f(\phi)$, provided that $f'^2\gg f^M,f$.

In the special case $M=2$ one sees
that $\ddot\phi=\dot\phi=0$ is a solution of
(\ref{inse}). One has then an asymptotic deSitter inflation, analogously
to what occurs in NMC models. We will not give details of
this case, since the dynamics depends on
the behavior of $V$ and $f$ at small $\phi$, while here we are
interested mainly in the asymptotic properties.

Let us come to the third, more interesting, case, namely $M> 2$.
Now $f^M$ dominates in $A$ and $f^{M-1}$ in $B$. Eq. (\ref{inse}) then reads
\beq\label{scase}
\Box\phi+{f'\over f}\dot\phi^2 (M-1)+{\xi\over 4\mu}f'(M-2)=0\,.
\eeq
It worths remarking that the effective potential is
$U(\phi)=\xi (M-2) (4\mu)^{-1} f(\phi)$; the
coupling function $f(\phi)$ reveals itself as an effective potential
for the theory. Notice also that, contrary to the
previous case, the constant $\mu$ plays here an important role.
Let us put now
$f(\phi)=x^m$, where $x\equiv\phi^2$. It then follows
\beq\label{ics}
\Box x-{\dot x^2\over 2x}[1-2N-2m]+ {\xi N\over \mu}x^m=0\,,
\eeq
where $N=m(M-2)$,
for $x\ge 0$. To have a positive-definite potential derivative is
sufficient to have $M>2$ (provided that $\xi/\mu>0$) .

To find explicit solutions to Eq. (\ref{ics}) we need also
$H=H(\phi,\dot\phi)$. From Eq. (\ref{00con}), neglecting all terms higher
than first order and in the limit $\xi f\gg 1$, we get for
$M>2-1/m$ (i.e. $N>-1$)
\beq\label{h2}
[H^{(1)}]^2= {\lam\over 3|\xi|} x^{N-1}
\left[ {3\mu\over\xi} \dot x^2+x^{m+1}\right]\,.
\eeq
Let us put a trial solution $\dot x=b x^p$ in (\ref{ics}) and
(\ref{h2}). In the limit of large $x$ we find the  attractor
\beqa\label{attr}
p &=& (m-N)/2\,,\nonumber\\
b &=& -{N\over \sqrt{3}}\left({\xi^3\over\lam \mu}\right)^{1/2}\,.
\eeqa
It can be easily proved that this really is an attractor; a
graphic evidence is provided by our phase-space portraits,
Figs (\ref{f1},\ref{f2}).
Along the
solution (\ref{attr}) the term $\ddot x$ is negligible; this justifies
the second order approximation previously performed. Integrating
$\dot x=bx^p$ one gets
\beq\label{icst}
x=x_0(1+t/\tau)^{1/(1-p)}\,,
\eeq
where $\tau^{-1}=b(1-p)x_0^{p-1}$ and  $x(t=0)=x_0$. For
$p<1$ the time constant $\tau$ is negative; clearly our approximations
break down for $t\to \tau$. The inflationary regime we will find also
breaks for $t\to\tau$.
Along the attractor we have the Hubble function
\beq\label{hattr}
H=\left(\lam/3|\xi|\right)^{1/2} x^{(N+m)/2}\,,
\eeq
and the cosmic scale factor
\beq\label{adit}
a(t)=a_0\exp\left\{E \left[1- (1+t/\tau)^k\right]\right\}\,,
\eeq
where
\beq\label{ef}
E={3(|\mu/\xi|) x_0^{N+1}\over N(N+1)}\,,\qquad
k={2(N+1)\over 2+N-m}\,.
\eeq
The scale factor expansion
is {\it always} inflationary for $t< \tau$.
For $t\ll \tau$ the expansion follows the deSitter law
$a\sim \exp [(E|k/\tau|)t]$, ($k/\tau$ is always negative).
The constant $E$ sets the total number of $e$-foldings of the
inflationary stage. If, for instance,
$\xi$ and $\mu$ are approximatively equal,
values of $\phi_0=x_0^{1/2}$ around unity in Planck units are
required to have $E>60$, the same requirement commonly found
in chaotic inflation.
This kind of behavior  is often called quasi-deSitter inflation.
The narrow
range in which viable inflation is found in NMC theories\bib{ABO},
 $2\le M<2 +\sqrt{3}$, is here expanded to $M\ge 2$, {\it
without upper bound}. The effect of including a derivative coupling is then
to substantially enlarge the class of viable inflationary models. This is
the main result of this paper.

It can be checked
that $H(t)$ and $x(t)$ decrease as the time increases.
As $|\phi|$ decreases, the solution crosses to
the NMC region, where the derivative coupling is negligible compared
to the non-derivative one.
In particular, the attractor enters the NMC region for
\beq\label{crit}
x < x_1= \left[{3\xi^2 m^2\over 2 |\mu|\lam}\right]^{1/(N+1)}\,.
\eeq
As $|\phi|$ further decreases, $\xi f(\phi)$ becomes smaller than unity
for  $x<x_2=\xi^{-1/m}$  and the attractor enters the minimally
coupled central region. For instance, in a model in which
$|\mu|\lam=3/2$ and $M=3, m=1$, one has $x_1=\xi$ and $x_2=1/\xi$.
If $x_2>x_1$ the NMC phase disappears.

In Figs. (\ref{f1},\ref{f2}) we show the numerical phase space of the
model for some parameter values. The plots are obtained integrating
Eq. (\ref{inse}) for several initial conditions, and then performing
a Poincar\'e projection onto the unitary circle. For any trajectory
we can identify four
stages, or cosmological epochs. All trajectories start
ideally from the initial singularity at $\phi,\dot\phi\to \infty$, and
they fall after a short transient on the inflationary attractors. In this
first stage the higher-order terms that we are
neglecting in this section can be important. The second stage
is represented by the outer part of the attractors, where  the $\mu$-terms
are dominant; now the trajectories follow Eq. (\ref{attr}), and the cosmic
expansion is given by (\ref{adit}). The third stage begins
when $x< x_1$, where the attractors
enter the NMC region. Now the behavior will be as described in
the introduction, with the  power-law expansion (\ref{NMC}) (inflationary
if $2\le M<2+\sqrt{3}$). Finally, the coupling terms become negligible
(for $x<x_2$) and the theory reduces to ordinary gravity.
The details of this last stage depend on the specific form of $V(\phi)$ for
small $\phi$, but eventually (perhaps after
a last quasi-deSitter inflationary episode) the mass term will dominate, and
a series of damped oscillations around the potential ground state
will occur. A Friedmannian expansion then takes place.
 The basin of attraction of the attractors
extends to almost all of the phase space, and all
initial conditions lead to the central Friedmann region. Since the
higher-order terms do not introduce new critical points in the model, the
qualitative picture given here should be of general validity.
Notice that Fig. (\ref{f2}) displays a model with $M=4$,
outside the successful range in
NMC models, but inflationary in the NMDC model.

\section{\normalsize\bf Conclusions}
We have shown that nonminimal derivative couplings are an
interesting source of new cosmological dynamics. Generally
speaking, their presence allows inflationary attractors where
otherwise they would be absent. The quite narrow range
of parameters that allows a viable inflation in NMC theories is
here expanded to a semi-infinite range. Let us remark that the succession
of different inflationary epochs may be used to break the
scale-invariance of inflationary perturbation spectra\bib{ST2}. In the
model effectively considered, three inflationary phases (quasi-deSitter,
power-law, quasi-deSitter) may occur, depending on the
parameter. This realizes a scenario of double (or triple)
inflation\bib{ST2} without the need to introduce more fields.

Before concluding, let us comment on another possible
consequence of a NMDC cosmology.
The inflationary attractors found in the previous section can be
used to implement a model of primordial phase transition
along the scheme of extended inflation. Suppose a second scalar field
$\psi$ with double-well self-interaction potential $V_2(\psi)$ is
present in the theory. During the inflationary stage driven by $\phi$
the new field may perform a first-order phase transition from
the more energetic false-vacuum state to the less energetic true
vacuum. The transition proceeds through quantum
and thermal nucleation of true-vacuum bubbles, whose surface energy
density contains the ``latent heat'' of the process. The bubbles
grow converting the background false-vacuum energy density into surface
``kinetic energy''
and eventually coalesce and percolate, unless the false-vacuum inflationary
expansion is too rapid. If $\Gam$ is the nucleation rate, the
(not normalized) probability
that a point is still in the false-vacuum state at the time $t$ is\bib{WEIN}
\beq\label{prob}
p_{FV}(t)=a^3(t) e^{-I(t)}\,,
\eeq
where
\beq\label{I}
I(t)={4\pi/3}\int_0^t dt' \Gam a^3(t') \left[\int_{t'}^t {du\over
a(u)}\right]^3
\,,\eeq
where $t=0$ is the instant in which the nucleation begins.
(Let us remark that $p_{FV}$ should be calculated in an Einstein frame;
for what concerns the following discussion this precisation is of
secundary importance.)
If $p_{FV}\to 0$ the transition is completed. It is well known
that in old inflation, where the expansion is exponential,
the transition can never be completed; extended inflation cures this
``graceful exit'' problem by slowing down the inflation to a power-law.
In our model the scale factor expands according to Eq. (\ref{adit})
during the NMDC phase (the second stage) and according to Eq. (\ref{NMC})
during the NMC phase (the third stage). It is the
latter phase to be crucial.
During the quasi-deSitter NMDC expansion the false vacuum grows so fast
that the bubbles nucleated are rapidly diluted before they can coalesce,
unless $\Gam$ is unrealistically large. This is the same problem one has
in old inflation. However, during the subsequent NMC epoch the false vacuum
slows down to a power-law expansion (it does not matter here whether
inflationary
or not) and the phase transition can find a natural exit, just as one
has in extended inflation. One should also ensure that the possible last
inflationary phase in the minimally coupled regime, which depends on the
details for $|\phi|\to 0$ of $V(\phi)$, be negligible. In this scenario,
the transition occurs almost entirely during the NMC epoch, when the
scale factor expands slowly  enough. Also the bubble nucleation will take place
mostly in the NMC epoch. Depending on the relative duration of the
NMDC inflationary phase and the NMC phase one
can have different bubble spectra. If the quasi-deSitter
NMDC phase lasts for a large
fraction of the 60 or so $e$-foldings one needs for
the inflation to be successful,
the bubbles produced during the subsequent power-law NMC phase are very small,
they rapidly thermalizes (perhaps with a production of gravitational waves
\bib{KTW})
 and do not affect neither the large scale structure, nor the cosmic
microwave background\bib{TUR}. If, on the contrary, the NMC phase takes, say,
55 $e$-foldings, then the bubbles would be today very large and hardly
thermalized. The cosmic microwave background measurements
and the primordial nucleosynthesis would put very
stringent limits on the model, but the intriguing possibility that the
primordial bubbles do
participate in the large-scale structure formation arises\bib{LA}.

\vspace{.2in}
\centerline  {ACKNOWLEDGMENTS}

The author thanks A. A. Starobinsky, H.-J. Schmidt
and G. Pollifrone for thoughtful comments on the manuscript.

\newpage
\centerline  {\bf Figure Caption}
\begin{enumerate}
\item\label{f1}
Poincar\'e phase space for a model with $m=1, M=3, \xi=\mu=-1, \lam=3/2$.
The inflationary attractors are clearly seen. The trajectories start
at $t=0$ from the ``north'' and ``south'' poles and rapidly reach the
attractors. Eventually, all trajectories fall onto the global stability
point at $\dot\phi=\phi=0$,
after a number of damped oscillations. The discrete symmetry $\phi,\dot\phi\to
-\phi,-\dot\phi$, also evident in the Lagrangian, shows up.
\item\label{f2}
Poincar\'e phase space for a model similar to the precedent,
but with $M=4$. In a NMC theory this model
would not be inflationary.
\end {enumerate}

\vspace{.5in}
\centerline{\bf References}
\frenchspacing
\begin{enumerate}

\item\label{LIN}
	A. Linde, \rpl 129B (1983) 177;
	 \rnp B216 (1983) 421 (1983).
\item\label{LM}
	F. Lucchin and S. Matarrese, \rpr D32 (1985) 1316.
\item\label{WM}
R. Wald, \rpr D28 (1983) 2118;
K.-I. Maeda, \rpr D37 (1988) 858.
\item\label{KFM}
	U.  Kasper, Nuovo Cim.   B103 (1989) 291;
	T. Futamase and K. Maeda,\rpr D39 (1989) 399;
	L. Amendola, M. Litterio and F. Occhionero,
	\rijmp A 5 (1990) 3861.

\item\label{JBD}
	P. A. M. Dirac, Proc. R. Soc.   A338 (1974) 439;
	 P. Jordan, Schwerkraft und Weltall (Wieweg, Braunschweig, 1955);
	C. Brans and R. H. Dicke \rpr,   124 (1961) 925.
\item\label{BD}
N. D. Birrell and P. C. W. Davies, Quantum Fields in
Curved Spaces (Cambridge Univ. Press, Cambridge, 1982).
\item\label{ZA}
	A. Zee, \rprl,   42 (1979) 417;
	F. S. Accetta, D. J. Zoller and M. S. Turner, \rpr D31 (1985)
	3046.
\item\label{KT}
E. W. Kolb and M. S. Turner,  The
Early Universe (Addison-Wesley, Menlo Park, CA, 1990).
\item\label{SS}
S. Randjbar-Daemi, A. Salam, and J. Strathdee, \rpl   B135 (1984) 388;
K. Maeda, \rcqg   3 (1986) 233;
T. Applequist and A. Chodos, \rprl   50 (1983) 141.

\item\label{MM}
M. Morikawa, \rapjl, 362 (1990) L37.
\item\label{YOK}
J. Yokohama, \rpl, B212 (1988) 273.
\item\label{BM}
J. Bekenstein and M. Milgrom, \rapj, 286 (1984) 7.
\item\label{LS}
	D.  La and P. J. Steinhardt, \rprl,   62 (1989) 376;
	D. La and P. J. Steinhardt, \rpl   B220 (1989) 375;
	F. S. Accetta and J. J. Trester, \rpr D  39 (1989) 2854;
	A. Linde, CERN-TH-5806/90 preprint (1990).
\item\label{SA}
	P. J. Steinhardt and F. S. Accetta,
	\rprl   64 (1990) 2740.
\item\label{HK}
        R. Holman, E. W. Kolb, and
	Y. Wang, \rprl   65 (1990) 17;
	R. Holman, E. W. Kolb, S. L. Vadas and Y. Wang, \rpr   D43 (1991)
	995; D43 (1991) 3833; Y. Wang, \rpr D  42 (1990) 2541.
\item\label{ACLO}
Y. Wang, \rpr   D42 (1990) 2541;
D. La, \rpr   D44 (1991) 1680;
L. Amendola, S. Capozziello, M. Litterio and F. Occhionero, \rpr
D45 (1992) 417.
\item\label{DGG}
T. Damour, G. W. Gibbons and C. Gundlach, \rprl   64 (1990) 123;
G. Piccinelli, F. Lucchin and S. Matarrese, \rpl   B277 (1992) 58.
\item\label{ABO}
L. Amendola, D. Bellisai and F. Occhionero, in preparation.

\item\label{W}
	B. Whitt, \rpl   145B (1984) 176;
	J. D. Barrow and Cotsakis S.,\rpl,   214B (1988) 515;
	K.-I. Maeda, \rpr D39 (1989) 3159;
	H.-J. Schmidt, Class. Q. Grav.   7 (1990) 1023.
\item\label{MFF}
G. Magnano, M. Ferraris and M. Francaviglia, \rgrg, 19 (1987) 465.
\item\label{ST2}
L. A. Kofman, A. Linde and A. Starobinsky, \rpl B157 (1985) 361;
L. Amendola, F. Occhionero and D. Saez, \rapj 349 (1990) 399.
\item\label{WEIN}
A. H. Guth and E. J. Weinberg, \rpr D23 (1981) 876;
\rnp B212 (1983) 321.
\item\label{KTW}
M. S. Turner and F.  Wilczek \rprl 66 (1991) 5.
\item\label{TUR}
M. S. Turner, Fermilab-Conf-91/220-A, preprint (1991);
M. S. Turner, E. J. Weinberg and L. M. Widrow, Fermilab-Pub-91/334-A,
preprint (1992).
\item\label{LA}
D. La, \rpl B265 (1991) 232.

\end{enumerate}

\end{document}